\documentclass{pasj00}

\Received{$\langle$reception date$\rangle$}
\Accepted{$\langle$acception date$\rangle$}
\Published{$\langle$publication date$\rangle$}

\SetRunningHead{Astronomical Society of Japan}{Usage of \texttt{pasj00.cls}}

\usepackage{times}

\begin{document}

\title{The Statistical Analyses of the White-Light Flares:\\
Two Main Results About Flare Behaviours}

\author{Dal, H. A.}
\affil{Department of Astronomy and Space Sciences, University of Ege, Bornova, 35100 ~\.{I}zmir, Turkey}

\email{ali.dal@ege.edu.tr}

\KeyWords{methods: data analysis --- methods: statistical --- stars: activity --- stars: flare --- stars: individual(CR Dra)}

\maketitle

\begin{abstract}

We present two main results, based on the models and the statistical analyses of 1672 U-band flares. We also discuss the behaviours of the white-light flares. In addition, the parameters of the flares detected from two years of observations on CR Dra are presented. By comparing with the flare parameters obtained from other UV Ceti type stars, we examine the behaviour of optical flare processes along the spectral types. Moreover, we aimed, using large white-light flare data,to analyse the flare time-scales in respect to some results obtained from the X-ray observations. Using the SPSS V17.0 and the GraphPad Prism V5.02 software, the flares detected from CR Dra were modelled with the OPEA function and analysed with t-Test method to compare similar flare events in other stars. In addition, using some regression calculations in order to derive the best histograms, the time-scales of the white-light flares were analysed. Firstly, CR Dra flares have revealed that the white-light flares behave in a similar way as their counterparts observed in X-rays. As seen in X-ray observations, the electron density seems to be a dominant parameter in white-light flare process, too. Secondly, the distributions of the flare time-scales demonstrate that the number of observed flares gets a maximum value in some particular ratios, which are 0.5 or its multiples, and especially positive integers. The thermal processes might be dominant for these white-light flares, while non-thermal processes might be dominant in the others. To reach better results for the behaviour of the white-light flare process along the spectral types, much more stars in a wide spectral range, from the spectral type dK5e to dM6e, must be observed in the white-light flare patrols.

\end{abstract}

\section{Introduction}

The flare process has not been perfectly understood yet, although several studies on this subject have been carried out since the first flare was detected on the Sun by R. C. Carrington and R. Hodgson on September 1, 1859 \citep{Car59, Hod59}. The flare activity of dMe stars is modelled based on the processes of Solar Flare Event \citep{Ger05, Ben10}. To understand flare event of dMe stars, the levels of the flare energy have been looked over in many studies. \citet{Ger83} demonstrated that there are some differences between energy levels of different stars and star groups. They demonstrated that the flare energy levels of the Orion association stars are located at the highest levels. The Pleiades stars are located just below the Orion association stars. Finally, the stars in the Galactic field are located below all of them. The levels seem to be arranged according to the different ages. However, the differences among the different stars and star groups might be due to the saturation level of white-light flares detected from UV Ceti type stars. The X-ray and radio observations indicate that some parameters of magnetic activity can reach the saturation \citep{Ger05, Sku86, Vil86, Doy96, Gar08, Jef11}. \citet{Dal11a} recently revealed some clues for the saturation in the white-light flares. They modelled the distributions of flare-equivalent durations versus flare total duration. In the models, it is seen that flare-equivalent durations can not be higher than a specific value and it is no matter how long flare total duration is. According to \citet{Dal11a}, this level defined as the $Plateau$ parameter in the models is an indicator of the saturation level for the white-light flares in some respects.

To understand flare event of dMe stars, the flare time-scales are as important as emitting flare energy. There are some parameters (such as $B$, $n_{e}$ and geometry of flaring loop) to determine the border of the time-scales in the flare events. These parameters affect the time-scales of heating and cooling of the regions, where the flare events occur \citep{Tem01, Ree02, Ima03, Pan08}. According to \citet{Ima03}, the flare decay time ($\tau_{d}$) is firmly correlated with $B$ and the electron density ($n_{e}$), while the flare rise time ($\tau_{r}$) is proportional to a larger magnetic loop length ($\ell$) and smaller $B$ values. \citet{Ree02} concluded that the observed light curve is modelled with using multi-loop instead of a single-loop. However, some discrepancies were observed in the decay phases and especially toward the end of the flare light curves. In contrast to that assumed by \citet{Ima03}, \citet{Pan08} found that there is a sustained heating during the decay phases of the flares. Moreover, \citet{Fav05} demonstrated that short flares occur in a single loop, while longer flares are generally arcade flares and some heating is driving their decay phases. In addition, there are some relations between the flare rise and decay times. Using the event asymmetry index ($A_{ev}$), \citet{Tem01} statistically analysed solar $H_{\alpha}$ flares, and found some relations between flare decay and rise time-scales. \citet{Kay03} demonstrated that there is a linear correlation between flare rise and decay times. 

In the first part of this study, the flares of CR Dra were modelled, and some parameters were derived from the model. We compared CR Dra with other UV Ceti type stars in respect to these parameters. In this point, CD Dra takes an important place, because it fills a gap in the B-V range. CR Dra is a flare star \citep{Ger99}, which is classified as a double or multiple star in the SIMBAD database. The orbital period is 4.04 yr and the semimajor axis of the orbit is 0$\arcsec$.148. The total mass of the system is reported as a value between 1.0 $M_{\odot}$ and 1.8 $M_{\odot}$. The distance of the system is given between 17.4 $pc$ and 20.7 $pc$ \citep{Tam08}. CR Dra is a metal-rich stars and a member of the old disk population in the galaxy \citep{Sta86, Vee74}. The photometric magnitude difference between the components of the system was measured as $1^{m}.8$ in V-band by \citet{Tam08}. The flare activity of CR Dra was discovered for the first time by \citet{Pet57}. Apart from the flare activity, there are some studies for the sinusoidal-like variations at out-of-flares, such as \citet{Mah91, Mah93, And79}. However, it is not clear that the star exhibits any rotational modulation. \citet{Mah91} found that CR Dra exhibits a sinusoidal-like variation with period of $16^{d}$ in V-band, while the period found from B-band is $5^{d}$. The author mentioned that the amplitude of the variation found in B-band was lower that the observational standard deviation. On the other hand, \citet{Mah93} found that the period of the sinusoidal-like variation at out-of-flare is $8^{d}.776$ in V-band.

In the second part of this study, a large flare data containing 1672 U-band flares, which were obtained in this study and collected from 42 different studies in the literature, were analysed to test whether any relation exists between the flare rise and decay times for the white-light flares, or not.

\section{Observations and Data}

\subsection{Observations}

Observations were acquired with the High-Speed Three-Channel Photometer (HSTCP) attached to the 48 cm Cassegrain-type telescope at Ege University Observatory. Observations were performed in two different manners. We used a tracking star in the second channel of the photometer and carried out the flare observations only in standard Johnson U-band with exposure times between 2 and 10 seconds and time resolution of 0.01 seconds. The second type of observations was used for determining whether there was any variation out-of-flare. We also observed CR Dra once or twice a night, when the star was close to the celestial meridian. These observations were made with the exposure time of 10 seconds for each band of standard Johnson BVR system. Considering the technical properties of the HSTCP given by \citet{Mei02} and following the procedures outlined by \citet{Kir06}, the mean average of the standard deviations of observation times was computed as 0.08 seconds for the U-band observations. It was found to be 0.20 seconds for the multi-band observations. The same comparison stars were used for all observations. The basic parameters of all the program stars (such as standard V magnitudes and B-V colours) are given in Table 1. Although the program and comparison stars are so close in the plane of the sky, differential extinction corrections were applied. The extinction coefficients were obtained from observations of the comparison stars on each night. Moreover, the comparison stars were observed with the standard stars in their vicinity and the making use of differential magnitudes, in the sense of variable minus comparison, were transformed to the standard system using the procedures outlined by \citet{Har62}. The standard stars are listed in the catalogue of \citet{Lan83}. Heliocentric corrections were applied to the observation times. The mean averages of the standard deviations are $0^{m}.015$, $0^{m}.009$, $0^{m}.007$ and $0^{m}.007$ for the observations acquired in standard Johnson UBVR bands, respectively. To compute the standard deviations of observations, we use the standard deviations of the reduced differential magnitudes in the sense comparisons (C1) minus check (C2) stars for each night. There is no variation in the standard brightness comparison stars. The differential magnitudes in the sense of comparison minus check stars were carefully checked for each night. The comparison and check stars were found to be constant in brightness during the period of observations.

In general, 20 U-band flares were detected in observations of CR Dra. The star was observed $5^{h}.38$ in 2005 and $29^{h}.10$ in 2007. All the flares were detected in U-band observations of 2007. Apart from flares, no variation was found in BVR bands out-of-flare activity. Some samples of detected flares are shown in Figure 1. Several parameters, such as the flare rise time, decay time, amplitude and equivalent duration, were computed from the light curve of each flare. The procedure used in the calculations is similar to the methods found in the literature \citep{Ger72, Mof74}. The procedure schema of calculations are shown in Figure 2 for two flares, which are seen in both top and bottom panels of Figure 1. In calculations, we firstly separated each flare light curve into three parts. One of them is the part indicating the quiescent level of the brightness before the first flare on each nigh. The brightness level without any variations (such as a flare or any oscillation) was taken as a quiescent level of the brightness of this star. To determine this level, we used the standard deviation of each observation point, considering the mean average of all the observation points until this last point. If the standard deviations of the following points get over the $3\sigma$ level, this point was taken as the beginning of a flare. The quiescent levels of each star were determined from all the observation points before the first flare on each night. It must be noted that, although a different quiescent level was found for each observing night, their levels are almost the same considering the $3\sigma$ levels of observing condition. As seen in Figure 2, we fitted this level with a linear function ($f_{1}(x)$ and $g_{1}(x)$), and then, using this linear function, we computed the flare equivalent duration, flare amplitude and all the flare time-scales (rise and decay times). The part of the light curve above the quiescent level was also separated into two sub-parts. First of them is the impulsive phase, in which the flare increases. Second one is the decay phase. The impulsive and decay phases were separated according to the maximum brightness observed in this part. It must be noted that some flares have a few peaks. In this case, the point of the first-highest peak was assumed as the flare maximum. To determine the flare time-scales, we fitted the impulsive and decay phases with the polynomial functions. In Figure 2, $f_{2}(x)$ and $g_{2}(x)$ are the fits of the impulsive phases for the given samples, while $f_{3}(x)$ and $g_{3}(x)$ are the fits of decay phases. The best polynomial functions were chosen according to the correlation coefficients ($r^{2}$) of fits. To determine the beginning and end of each flare, we computed the intersection points of the polynomial fits with the linear fit of the quiescent level and their standard deviations. In this study, the intersection points were taken as the beginning and end of each flare. The flare rise time ($\tau_{r}$) was taken the duration between the beginning and the flare maximum point. For example, the duration from the first intersection between $f_{1}(x)$ and $f_{2}(x)$ to the second one between $f_{2}(x)$ and $f_{3}(x)$ was taken as the flare rise time ($\tau_{r}$) for this flare. In the same way, the flare decay time ($\tau_{d}$) was taken the duration between the flare maximum point and the flare end. In Sample 1, the duration from the second intersection between $f_{2}(x)$ and $f_{3}(x)$ to the third one between $f_{3}(x)$ and $f_{1}(x)$ was taken as the flare decay time ($\tau_{d}$). The height of the observed-maximum points from the $f_{1}(x)$ linear fit was taken as the amplitudes of this flare. The same procedure was used for each flare, and the Maple 12 \citep{Mon08} software was performed in all calculations.

Apart from the flare time-scales and amplitudes, using Equations (1) and (2) taken from \citet{Ger72}, the equivalent durations and energies of all the flares were computed: 

\begin{center}
\begin{equation}
P = \int[(I_{flare}-I_{0})/I_{0}] dt
\end{equation}
\end{center}
where $I_{0}$ is the flux of the star in the observing band while in the quiet state, and $I_{flare}$ is the intensity at the moment of flare. For instance, the parameter $P$ is equal to about total area between the $f_{1}(x)$, $f_{2}(x)$ and $f_{3}(x)$ for Sample 1.

\begin{center}
\begin{equation}
E = P \times L
\end{equation}
\end{center}
where $E$ is the energy, $P$ is the flare-equivalent duration in the observing band, and $L$ is the intensity in the observing band while the star is in the quiet state.

For each observed flare, the HJD of flare maximum moment, flare rise ($\tau_{r}$) and decay time ($\tau_{d}$) (s), flare total durations (s), ratio of $\tau_{d}/\tau_{r}$, flare-equivalent duration (s), flare amplitude (mag), and their energies (erg) were calculated. Those parameters are given in Table 2.

\subsection{Data Used In the Analyses}

The number of flares used in the analyses are 1672 from which 20 flares are detected in the observations of CR Dra in this study. 534 flares were obtained from other stars of this project. The remaining 1118 flares were collected from 41 different studies in the literature. Thus, the data used in the analyses were contained the parameters of 15 different UV Ceti stars. In Table 3, the distribution of the data taken from the literature is listed versus references ordered by year.

In the analyses of flare activity, especially in the OPEA models, 554 flares obtained in this project were only used. This is because the analyses need to be comprised of parameters derived with the same method and from the flares detected with the same optical system. Otherwise, some artificial variations and differences can occur between the data sets. To avoid this problem, instead of all 1672 flares, just 554 flares detected in this project, which comprised parameters derived with the same method and the same optical system, were used. On the other hand, all 1672 flares were used in the analyses of the distribution of the flare numbers versus the ratio of flare decay time to rise time. Determining and computing of the time-scales of the rise and decay phases are almost the same in all the studies.

\section{Analyses}

\subsection{Flare Activity and the One-Phase Exponential Association Models}

The distributions of the equivalent durations in the logarithmic scale versus flare total durations were derived in order to test whether there are any upper limits for the distributions of the equivalent durations ($log P_{u}$) of the flares detected in the observations of CR Dra. Although the flare energy is generally used to analyse in the literature, the flare equivalent durations were used to analyse in this study. This is because of the luminosity parameter ($L$). As seen from Equations (2), the flare energy ($E$) depends on the luminosity parameter ($L$) with flare equivalent duration ($P$). However, the luminosity ($L$) is different for each star. Although there are small differences among the masses of M dwarfs, the luminosities of the two M dwarfs, whose masses are very close to each other, can be dramatically different due to their position in the H-R diagram. This means that the computed energies of flares are very different from each other, even if the light variations of the flares occurring on these two stars are absolutely the same. Because of this, the equivalent durations ($P$) were used instead of energy ($E$) in this study. The equivalent duration parameter ($P$) depends only on flare power.

In order to model this distribution, first of all, the best curve was estimated with SPSS V17.0 software \citep{Gre99} and GraphPad Prism V5.02 software \citep{Mot07}. The regression calculations showed that the Exponential function is the best model. Besides, the tests done with GraphPad Prism V5.02 software revealed that especially the One Phase Exponential Association function (hereafter OPEA) \citep{Mot07, Spa87} given by Equation (3) is the best exponential function among all others. The OPEA model of the distributions of the equivalent durations in the logarithmic scale versus flare total durations was derived with using the Least-Squares Method with GraphPad Prism V5.02 software.

\begin{center}
\begin{equation}
y~=~y_{0}~+~(Plateau~-~y_{0})~\times~(1~-~e^{-k~\times~x})
\end{equation}
\end{center}

There are some important parameters derived from this function, which have some clues about the flaring loop and the condition inside the loop. One of them is $y_{0}$, which is the lower limit of equivalent durations for observed flares in the logarithmic scale. In contrast to $y_{0}$, the parameter of $Plateau$ is the upper limit. The value of $y_{0}$ depends on the quality of observations as well as flare power, while the value of $Plateau$ depends only on power of flares. In fact, according to Equation (2), the value of $Plateau$ depends only on the energy of flares occurring on the star. The parameter $k$ in Equation (3) is a constant depending on the $x$ value, which is the total flare time. Apart from them, one of the most important parameter is the $Half-Life$ parameter. This one is half of the first x values, where the model reaches the plateau values for a star. The $Half-Life$ parameter was also computed from the derived model. The derived OPEA model is shown in top panel of Figure 3, and the parameters of the model are listed in Table 4. In the middle panel of Figure 3, the model derived for CR Dar is compared with the models derived for other six UV Ceti type stars \citep{Dal11a, Dal11b}. In the bottom panel of the figure, the $Plateau$ values are compared for these stars.

Considering the flares in the $Plateau$ phases of the OPEA model, the mean average of maximum flare-equivalent durations was computed to test whether the $Plateau$ value is statistically acceptable, or not. The Independent Samples t-Test (hereafter t-Test) \citep{Wal03, Daw04} was used for the calculation. Thus, the value of $Plateau$ was tested using another statistical method. Although the mean value computed by the t-Test is expected to be close to the $Plateau$ value of the OPEA model, it is clear that there can be some difference between these two values. This is because the OPEA model depends on all the distributions of the $x$ values from the beginning to end, while the mean value computed by the t-Test depends only on equivalent durations of flares in the $Plateau$ phase. The results obtained from the t-Test analyses are also listed in Table 4.

Examining the flares detected in U-band observations of CR Dra, it was computed that the maximum flare rise time obtained from these 20 flares is 1967 s, while the maximum flare total duration is 4955 s.

\subsection{Distribution of the Flare Numbers versus the Ratio of Flare Decay Time to Flare Rise Time}

Several parameters were computed from the light curves of all these flares detected in this project. Then, using some statistical methods, the relations among themselves of all the parameters were analysed with both SPSS V17.0 \citep{Gre99} and GraphPad Prism V5.02 software \citep{Mot07}. During the analyses of the distribution of flare amplitudes versus the ratio of flare decay time to rise time, a remarkable gathering was seen for some ratios of flare decay time to rise time, which values are several times of the values 0.5 and especially the positive integer numbers, such as 0.5, 1.0, 1.5, 2.0, 2.5, 3.0, 3.5, 4.0, 4.5, etc. These accumulations are seen in Figure 4. As seen from the figure, the ratios of flare decay times to rise times are especially positive integers for many flares with different amplitudes.

To examine this unusual unexpected case, first of all, the ratio of flare decay time to rise time were computed for each flare. The values of $\tau_{d}/\tau_{r}$ are listed for CR Dra flare in Table 2. Considering the errors listed in Table 2, the mean error of $\tau_{d}/\tau_{r}$ values was found to be $\pm$0.042 for the flares detected from CR Dra. However, in this project, 554 U-band flares are detected in total. Taking all 554 flares into account, the mean error of $\tau_{d}/\tau_{r}$ values was found to be $\pm$0.026. As also seen from Table 2, although there are some flares, whose $\tau_{d}/\tau_{r}$ errors are dramatically larger than the computed mean errors, the total number of these flares is small enough to enough to be neglected. Apart from the flare detected in this project, we also collected 1118 flares from the literature and combined them with 554 flares. The sources of 1118 flares are listed in Table 3. It must be noted that unfortunately we could not compute the errors of $\tau_{d}/\tau_{r}$ values for the flares taken from the literature due to absence of errors of $\tau_{r}$ and $\tau_{d}$ values in the sources. Secondly, the best histograms were determined using SPSS V17.0 and GraphPad Prism V5.02 software for these data. Table 5 shows the parameters use to produce the histograms. The analyses indicate that the best statistically acceptable interval ratio length should be 0.05 in the interval ratio length.

The derived numbers of the flares in intervals of 0.05 were plotted versus the ratio of flare decay time to rise time. Then, the histogram was derived for 554 flares detected in this project. The ratio distribution is shown in Figure 5, while the results of the analysis is listed in Table 5. As seen from the figure, the ratio is equal to 1.0 in 0.05 ratio length for 73 flares, and it is 2.0 for 39 flares and also 3.0 for 29 flares. In brief, the ratios of flare decay time to rise time are especially equal to 1.0 or its multiples as the positive integers for 204 flares among 554 flares. Besides, the ratio is 0.5 for 18 flares. In total, the ratios of flare decay times to rise times are equal to 0.5 or its multiples, such as 1.5, 2.5, 3.5, 4.5, etc. The incidence of the flares, whose ratio of their decay times to rise times is 1.0 or positive integers, is 36.82$\%$ over 554 flares detected in this project. It is 15.70$\%$ for the flares, whose ratios are 0.5 or its multiples, such as 1.5, 2.5, 3.5, 4.5, etc. Considering that the positive integers are acceptable as the multiples of 0.5, the incidence of the flares, whose ratios of their decay times to rise times are 0.5 or its multiples, such as 1.0, 1.5, 2.0, 2.5, 3.0, 3.5, 4.0, 4.5, etc., is 52.53$\%$ over 554 flares.

In order to test whether this remarkable result obtained from the data of this project might be occurred due to any selection effects, almost all the U-band flare data given in the literature were collected from 41 different studies. Thus, 1118 U-band flares were collected apart from 554 U-band flares of this project. In fact, there are more 170 flares given in literature, but the decay or/and rise times were not given. Because of this, the ratios of flare decay times to rise times were not computed for these ones, and they could not be used. Following the method mentioned above, the same examinations were done for these 1118 U-band flares. The result is seen in Figure 6, while the results of the analysis is listed in Table 5. According to the statistical analyses, the incidence of the flares, whose ratio of their decay times to rise times is 1.0 or positive integers, is 16.37$\%$ over 1118 flares collected from the literature. It is 8.23$\%$ for the flares, whose ratios are 0.5 or its multiples, such as 1.5, 2.5, 3.5, 4.5, etc. Considering that the positive integers are acceptable as the multiples of 0.5, the incidence of the flares, whose ratios of their decay times to rise times are 0.5 or its multiples, such as 1.0, 1.5, 2.0, 2.5, 3.0, 3.5, 4.0, 4.5, etc., is 24.60$\%$ over these 1118 flares.

Finally, both the data obtained in this project and the data collected from the literature were combined together in order to reach rather better result. In brief, a data set containing 1672 U-band flares in total was obtained. Following the same method, the distribution was derived and statistically examined. The result of the examinations is shown in Figure 7, while the results of the analysis is listed in Table 5. Consequently, considering 1583 U-band flares, the incidence of the flares, whose ratio of their decay times to rise times is 1.0 or positive integers, was found as 23.09$\%$. It was computed as 10.83$\%$ for the flares, whose ratios are 0.5 or its multiples, such as 1.5, 2.5, 3.5, 4.5, etc. Considering that the positive integers are acceptable as the multiples of 0.5, the incidence of the flares, whose ratios of their decay times to rise times are 0.5 or its multiples, such as 1.0, 1.5, 2.0, 2.5, 3.0, 3.5, 4.0, 4.5, etc., is 33.91$\%$ over these 1583 flares.

\section{Results and Discussion}

\subsection{Flare Activity and the One-Phase Exponential Association Models}

The distributions of flare-equivalent durations versus flare total duration were modelled by the OPEA function expressed by Equation (3) for CR Dra. As it is seen from Figure 3, this function demonstrates that the equivalent durations of the flares detected in the observations of CR Dra can not be higher than a specific value regardless of the length of the flare total duration. In order to test whether the $Plateau$ parameter derived from the model is statistically acceptable, using the t-Test, the mean average of equivalent durations was computed for the flares located in the $Plateau$ phase in the OPEA models. The mean average of equivalent durations was found to be close to the $Plateau$ values derived from the OPEA models. The $Plateau$ parameter derived from the OPEA model was identified as an indicator of the saturation level for the white-light flares by \citet{Dal11a}.

According to \citet{Dal11a}, two stars, EV Lac and EQ Peg, have almost different level of the $Plateau$ value. However, the observations of CR Dra have demonstrated that the variation of the $Plateau$ value is different from that shown by \citet{Dal11a}. In fact, the shape of the $Plateau$ value variation is remarkably different from the conclusions of \citet{Dal11a}. EV Lac and EQ Peg are not different from the others. CR Dra flares indicate that the variation of this parameter has a trend, and both EV Lac and EQ Peg are on the trend. The observations of CR Dra also demonstrated that white-light flare activity could be affected by some parameters of the flare processes as well as the flares detected in X-ray or radio observations. The figure taken from \citet{Kat87} demonstrated that electron density varies along B-V indexes \citep{Ger05}, and its variation trend is in contrast shape with that of the $Plateau$ values. The analyses of flares detected in X-ray have revealed some important points of the processes. Considering the parameters of the Solar Flare Event, the suspected parameters could be one or several of $\nu_{A}$, $B$, $R$ or $n_{e}$. The suspected parameters must be $B$ or/and $n_{e}$ in the atmosphere of the stars during the flare event. According to \citet{Kat87} it could be $n_{e}$. On the other hand, \citet{Doy96} suggested that it can be related to some radiative losses in the chromosphere instead of any saturation in the process. However, \citet{Gri83} demonstrated the effects of radiative losses in the chromosphere on the white-light photometry of the flares. According to \citet{Gri83}, the negative H opacity in the chromosphere causes the radiative losses, and these are seen as pre-flare dips in the light curves of the white-light flares.

The other parameters derived from the OPEA model are the flare time-scales. The maximum flare duration was computed as 4955 s, while the maximum flare rise time was computed as 1967 s from the detected flares of CR Dra. The $Half-Life$ parameter was found to be 191.40 s from CR Dra flares. All these time-scales were found to be in agreement with those found for other stars by \citet{Dal11a, Dal11b}. It is well known that the decay time-scales of the X-ray light curves is related to the length of the loop \citep{Hai83, VanA88, VanB88, Rea97, Rea98, Rea88}. According to the results found by \citet{Fav05}, short flare durations could be an indicator of single loop geometry, while long flare durations could reveal arcade flares. If the same case is valid for the white-light flares observed from UV Ceti type stars, the lengths of the flaring loops on the surface of CR Dra are generally larger than those on AD Leo, EV Lac, EQ Peg according to the time-scales obtained from CR Dra. However, the general sizes of the flaring loops are almost equal to loops occurring on V1005 Ori. In the general perspective, the decreasing of the observed maximum time-scales (the flare rise time and the flare total duration) of flares reveals that the lengths of the flaring loop are decreasing toward the later spectral types among UV Ceti type stars. On the other hand, the flare mechanism of CR Dra easily reaches the saturation in shorter time than the stars such as AD Leo, EV Lac, EQ Peg. In this point, CR Dra and V1005 Ori have a same nature.

It must be noted that it is a debated issue whether the white-light flares can be affected from the parameters such as $\nu_{A}$, $B$, $R$ and $n_{e}$, which generally affect the coronal loops and the flare emissions in the corona \citep{Ben10, Haw92, Abb99, Haw03, All06}. However, the results found in this study reveal that there are some similarities between the white-light flares and their counterparts observed in the X-ray or radio bands.

Consequently, the observations of CR Dra have demonstrated that the $Plateau$ values vary in a trend from earlier spectral types to the later. However, the trend is not in the shape previously offered by \citet{Dal11a}. Although it is a debated issue, the shape of the trend in the $Plateau$ variation versus B-V indexes is in agreement with the trend in the distribution of $n_{e}$ versus B-V indexes. However, much more stars between the spectral type dK5e and dM6e should be observed in order to reach more reliable result. Especially, more stars around CR Dra should be observed.

\subsection{Distribution of the Flares Numbers versus the Ratio of Flare Decay Time to Flare Rise Time}

The most remarkable result was found in the analyses of the distribution flare amplitudes versus the ratios of decay times to rise times. As it can be seen from Figure 4, although the error bars of some flares are a bit large, the flares are gathering in the specific ratios. According to the distribution seen in the figure, many flares seem to prefer the value of 0.5 or its multiples, such as 1.0, 1.5, 2.0, 2.5, 3.0, 3.5, 4.0, 4.5, etc., in the ratios of decay times to rise times. To test whether these accumulations in these ratios are statistically meaningful, the histograms were derived and analysed to find statistically the best one. The derived best histograms are shown in Figures 5, 6 and 7. As seen from the figures, the statistical analyses indicated that the best width of bars is 0.05 for histograms. The mean error is found to be $\pm$0.042 for $\tau_{d}/\tau_{r}$ of the flares detected from CR Dra. Although the error is close to the estimated width value of bars, the mean error is lower. In addition, when the mean error computed over all 554 flares is $\pm$0.026. In this case, the histograms are statistically acceptable.

The histograms demonstrated some definite results. First of all, most of the flares prefer the value of 1.0 or its multiples as positive integers. The incidence of them over 1672 flares is 23.09$\%$. The incidence of the flares, which prefer the value of 0.5 or its multiples, such as 1.5, 2.5, 3.5, 4.5, etc., is 10.83$\%$. The positive integers can be taken as the multiples of the value 0.5. In this case, the incidence of the flares, whose ratios of the decay time to rise time are 0.5 and its multiples, reach the value of 33.91$\%$ over 1672 U-band flares. However, as it is stated in the Section 3.2, the incidences are found a bit different from the analysis of each data set. This must be because of some small differences in the methods used to compute the flare time-scales, such as decay or rise times. In this project, all the parameters were computed with considering the quiescent level of the brightness of a star. These levels were computed from the part of the observations without any flares or any variations for each night separately. On the other hand, in the literature, the first points of the flare beginnings were taken as a quiescent level of the brightness of observed star in some studies. Unfortunately, the first point of the flares does not always indicate the quiescent level of the brightness. Taking the first points of the flare beginnings as a quiescent level comes some difference to computed decay and rise times. Using the data collected from different studies might cause a bit different incidences.

In the case of many flares, preferring the ratios as 0.5 and its multiples might be related with some parameters in the flare event process, which affect the flare time-scales such as the flare rise or decay times. The flare decay time ($\tau_{d}$) is firmly correlated with $B$ and $n_{e}$, while the flare rise time ($\tau_{r}$) is proportional to a larger $\ell$ and smaller $B$ values \citep{Tem01, Ree02, Ima03, Pan08}. The magnetic loop length ($\ell$) also depends on plasma electron density ($n_{e}$) and plasma temperature \citep{Yok98, Shi99, Shi02, Yam02}. As it is seen, the preferred ratios of the decay times to rise times is mainly related with the electron density ($n_{e}$) in the flaring loop and quite a bit with magnetic field strength ($B$) of the loop. In fact, with some assumptions, \citet{Ima03} has shown the relations between flare time-scales (flare decay and rise times) and plasma electron density ($n_{e}$) and the reconnection factor ($M_{A}$) with Equations (A5) and (A8) given by them.

It is seen from Figures 5, 6 and 7, there are two type flares. One of them is the flares discussed above, which are shown by filled and open circles in the figures. Other type flares are shown by small points in the figures, which are neither the multiples of 1.0 nor the multiples of 0.5. These two types must be the slow and fast flare types \citep{Gur88}. \citet{Gur88} indicated that thermal processes are dominant in the processes of slow flares, which are 95$\%$ of all flares observed in UV Ceti type stars. Non-thermal processes are dominant in the processes of fast flares. In fact, the $\tau_{r}$ and $\tau_{d}$ parameters of the fast flares can take random values according to themselves \citep{Gur88, Dal10}. Thus, the small points, whose ratios are neither the multiples of 1.0 nor the multiples of 0.5, must indicate the fast flares.

Consequently, the distribution of the number of flares versus the ratio demonstrates that (1) the number of observed flares gets maximum in the ratio, which are 0.5 or its multiples, and especially positive integers. As a result of this, $\tau_{d}$/$\tau_{r}$ is usually equal to 0.5 or its multiples, and mostly a positive integer. (2) The flare numbers are dramatically decreasing toward the larger values of the ratio. (3) Some flares do not prefer specific values of the ratio, thus their ratios of the decay times to rise times are neither the multiples of 1.0 nor the multiples of 0.5. (4) The thermal processes might be dominant in the processes of the flares preferring specific ratios, while non-thermal processes might be dominant in the others.

\section*{Acknowledgments}
The author acknowledges the generous observing time awarded to the Ege University Observatory. I also thank the referee for useful comments that have contributed to the improvement of the paper.

\clearpage

\begin{figure*}
\hspace{6.2 cm}
\FigureFile(155mm,60mm){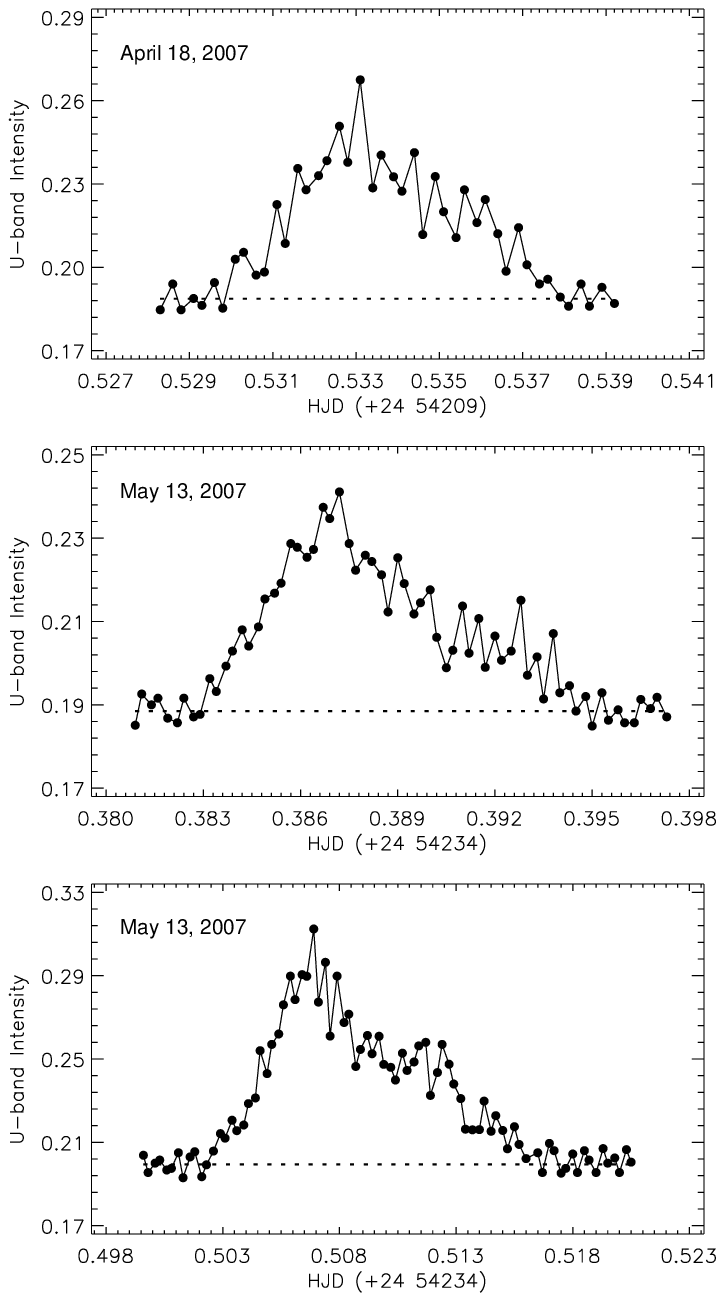}
\vspace{0.7 cm}
\caption{Three samples for the observed flare light curves in U-band. The flare shown in top panel was observed in April 18, 2007, while flares shown in middle and bottom panels were detected in May 13, 2007. In the figures, filled circle represents observations, while dashed line represents the quiescent level of the brightness determined for each night.\label{Fig.1.}}
\end{figure*}

\begin{figure*}
\hspace{1.2 cm}
\FigureFile(160mm,60mm){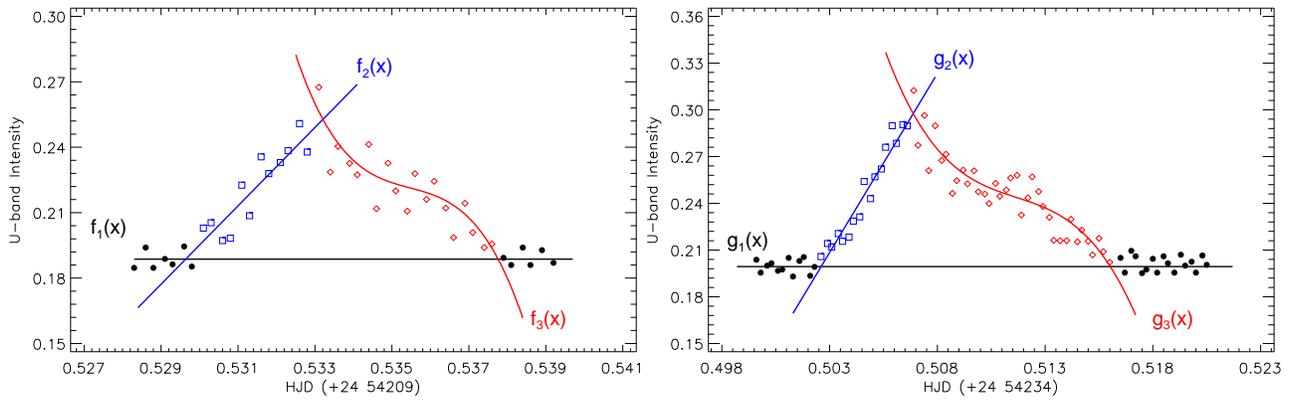}
\vspace{0.7 cm}
\caption{Two samples for computing of the flare parameters. In both figures, the filled-black circles represent observations in the quiescent level of the brightness. The blue-squares represent the impulsive phase of the flares, while the red-diamonds represent the main (decay) phases of the flares. The black lines ($f_{1}(x)$ and $g_{1}(x)$) represent the linear fits of the observations in the quiescent level, while the blue lines ($f_{2}(x)$ and $g_{2}(x)$) represent the linear fits of the observations in the impulsive phase. The red lines ($f_{3}(x)$ and $g_{3}(x)$) represent the polynomial fits of the decay phases of the flares.\label{Fig.2.}}
\end{figure*}

\begin{figure*}
\hspace{6.2 cm}
\FigureFile(170mm,60mm){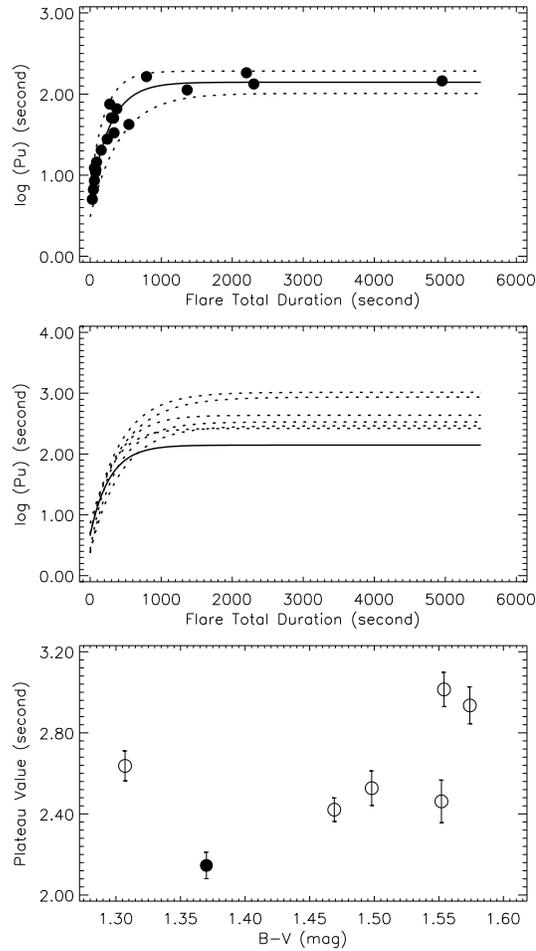}
\vspace{0.7 cm}
\caption{Comparison of CD Dra's The OPEA with other stars. Top panel: The OPEA model derived from the distributions of flare-equivalent duration on a logarithmic scale vs. flare total duration for CR Dra. Filled circles represent equivalent durations computed from flares detected from CR Dra. The line represents the model identified with Equation (3) computed using the least-squares method. The dotted lines represent 95$\%$ confidence intervals for the model. Middle panel: The OPEA model (solid line) of CR Dra is compared with the models (dotted line) derived from observations on other 6 stars reported by \citet{Dal11a, Dal11b}. Bottom panel: The $Plateau$ parameters vs. the B-V index of the CR Dra and other 6 stars taken from \citet{Dal11a, Dal11b} are shown. In the panel, open symbols represent the parameters taken from \citet{Dal11a, Dal11b}, while filled circle represents the parameter of CR Dra.\label{Fig.3.}}
\end{figure*}

\begin{figure*}
\hspace{1.5 cm}
\FigureFile(160mm,60mm){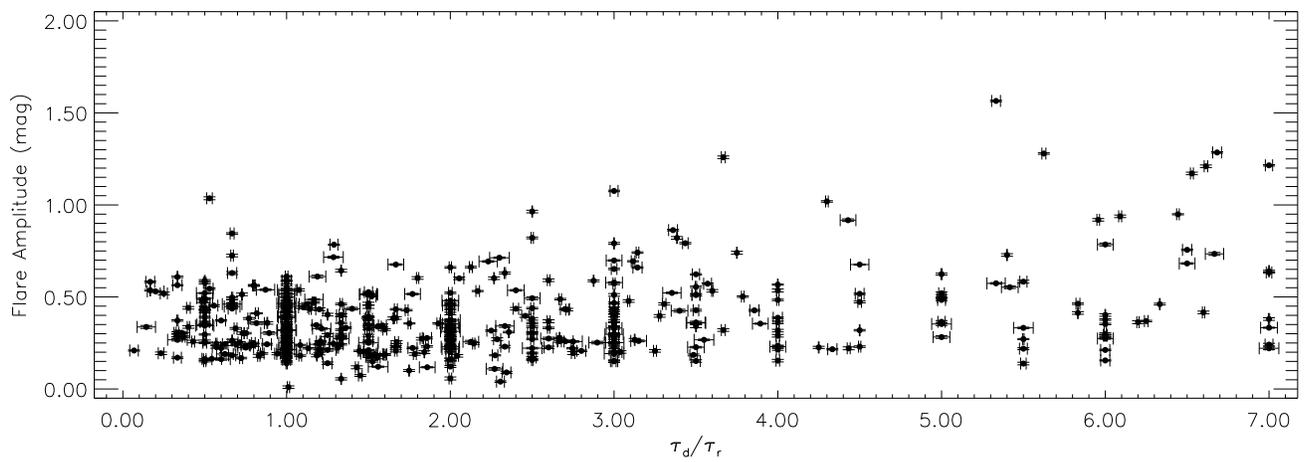}
\vspace{0.7 cm}
\caption{A sample for the distribution of the flare amplitudes versus the ratio of flare decay time to flare rise time. Instead of any amplitude variation in a regular trend, there is a remarkable gathering in some particular ratios of flare decay time to rise time.\label{Fig.4.}}
\end{figure*}

\begin{figure*}
\hspace{1.5 cm}
\FigureFile(155mm,60mm){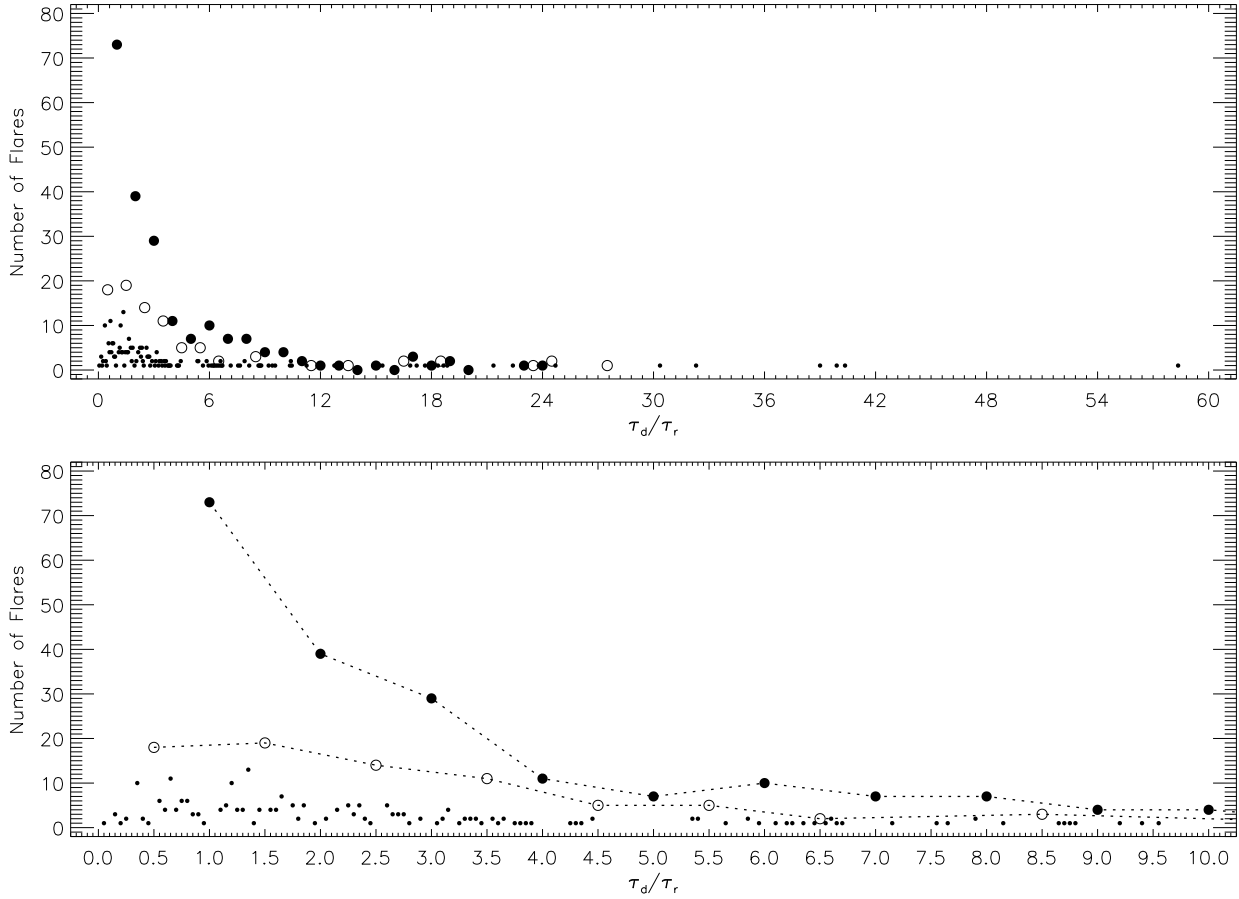}
\vspace{0.7 cm}
\caption{The distribution of the number of the flares derived in intervals of 0.05 ratio length versus the ratio of flare decay time to rise time. Top panel: the distribution is shown for ratios between 0.00 and 58.35. All the flares seen in this figure were obtained in this project. The value of 58.35 is the largest ratio obtained from the flares detected in this work. Bottom panel: the distribution is shown for the interval between 0.00 and 10.00 for clarity. In both panels, the filled circles represent the flares, whose ratios of flare decay times to rise times are equal to 1.0 or its multiples as the positive integers. The open circles represent the flares, whose ratio is 0.5 or its multiples, such as 1.5, 2.5, 3.5, 4.5, etc. The small points represent the flares with other ratios, which are neither the multiples of 1.0 nor the multiples of 0.5.\label{Fig.5.}}
\end{figure*}

\begin{figure*}
\hspace{1.5 cm}
\FigureFile(155mm,60mm){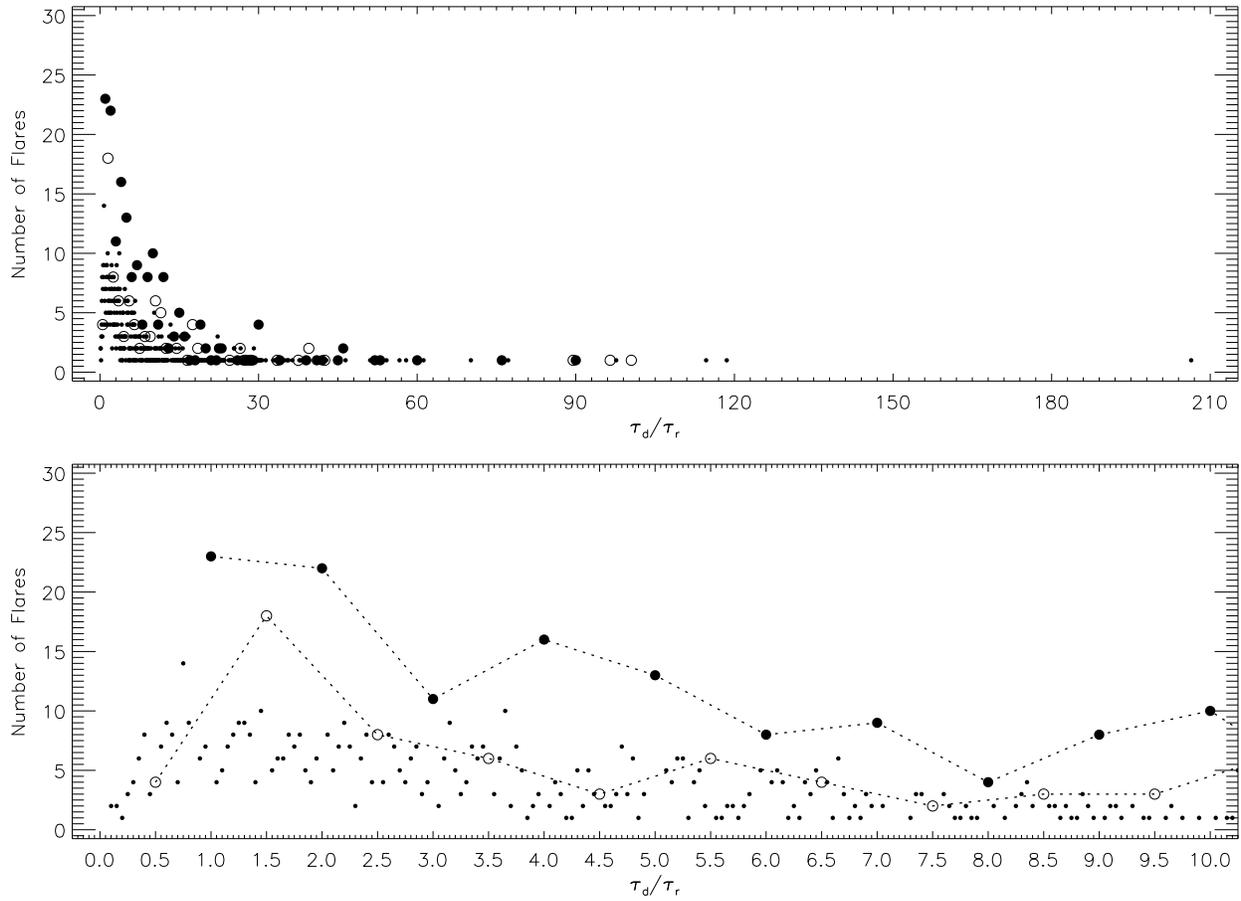}
\vspace{0.7 cm}
\caption{The distribution of the number of the flares derived in intervals of 0.05 ratio length versus the ratio of flare decay time to rise time. Top panel: the distribution is shown for all ratios from 0.00 to 206.40. All the flares seen in this figure were collected from the literature. The value of 206.40 is the largest ratio obtained from these flares. Bottom panel: the distribution is shown for the interval between 0.00 and 10.00 for clarity. In both panels, all the symbols used in this figure are the same with the symbols used in Figure 4.\label{Fig.6.}}
\end{figure*}

\begin{figure*}
\hspace{1.5 cm}
\FigureFile(155mm,60mm){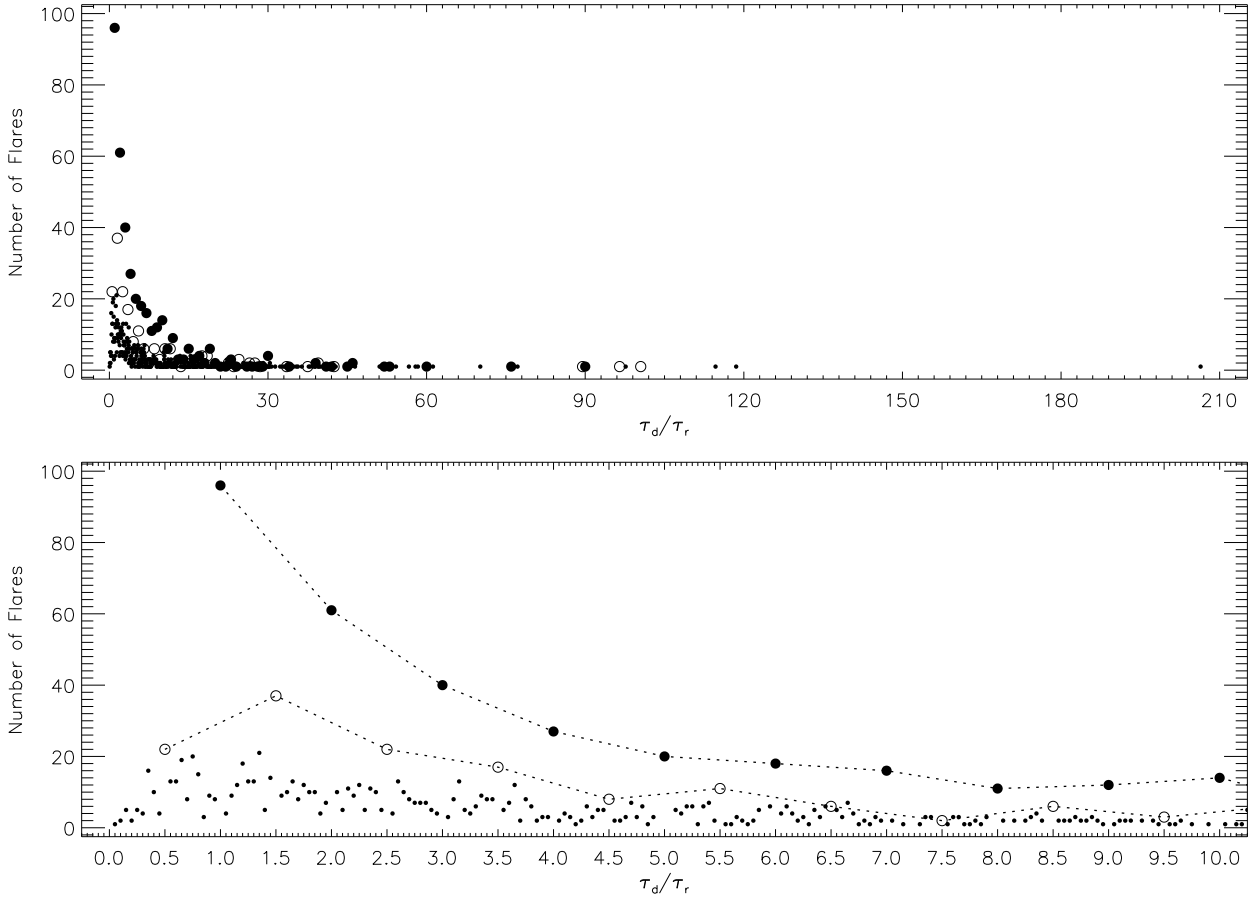}
\vspace{0.7 cm}
\caption{The distribution of the number of the flares derived in intervals of 0.05 ratio length versus the ratio of flare decay time to rise time. Top panel: the distribution is shown for all ratios from 0.00 to 206.40. The data plotted in the figure are a combination of data from this work and those taken from the literature. The value of 2006.40 is seen to be the largest ratio obtained from all the flares. Bottom panel: the distribution is shown for the interval between 0.00 and 10.00 for clarity. In both panels, all the symbols used in this figure are the same with the symbols used in Figure 4.\label{Fig.7.}}
\end{figure*}

\clearpage

\begin{table*}
\begin{center}
\caption{Basic parameters for the targets studied, comparison (C1) and check (C2) stars.\label{tbl-1}}
\begin{tabular}{lcc}
\hline\hline
\textbf{Program Stars}	&	\textbf{V (mag)}	&	\textbf{B-V (mag)}	\\
\hline				
\textbf{CR Dra}	&	9.997	&	1.370	\\
\textbf{C1: HD 238545}	&	10.195	&	0.582	\\
\textbf{C2: HD 238552}	&	10.123	&	1.021	\\
\hline
\end{tabular}
\end{center}
\end{table*}

\begin{table*}
\begin{center}
\caption{All the parameters computed for each flare detected in flare patrols of CR Dra in this study. All the flares were obtained in the U-band.\label{tbl-2}}
\begin{tabular}{crrrrrrr}
\hline\hline
HJD of Flare	&	Rise	&	Decay	&	Total	&		& Equivalent	&	Flare	&	Flare	\\
Maximum	&	Time	&	Time	&	Duration	&	$\tau_{d}/\tau_{r}$	& Duration	&	Amplitude	&	Energy	\\
(+24 00000) 	&	(s) 	&	(s) 	&	(s) 	&		& (s) 	&	(mag) 	&	(erg) \\
\hline
54208.47638 &  60$\pm$0.5  &  96$\pm$0.6 &  156$\pm$0.8  & 1.60$\pm$0.03 &  20.3403$\pm$0.0007  &  0.188$\pm$0.003  &  $\sim$ 9.5898 E+30 \\
54208.48038 &  110$\pm$1.1  &  2090$\pm$2.1  &  2200$\pm$2.4 & 19.00$\pm$0.2 &  183.0304$\pm$0.0024  &  0.518$\pm$0.008  &  $\sim$ 2.4743 E+32 \\
54209.53320 &  305$\pm$1.0  &  394$\pm$2.3  &  699$\pm$2.5  & 1.29$\pm$0.06 &  75.2661$\pm$0.0006  &  0.270$\pm$0.004  &  $\sim$ 3.5485 E+31 \\
54210.53322 &  204$\pm$0.5  &  341$\pm$1.1  &  545$\pm$1.2 & 1.67$\pm$0.01 &  42.2318$\pm$0.0005  &  0.255$\pm$0.004  &  $\sim$ 1.9911 E+31 \\
54210.54189 &  48$\pm$0.6  &  288$\pm$1.5  &  336$\pm$1.6 & 6.00$\pm$0.10 &  33.3097$\pm$0.0012  &  0.299$\pm$0.004  &  $\sim$ 1.5704 E+31 \\
54211.32701 &  48$\pm$0.6  &  33$\pm$0.4  &  81$\pm$0.7  & 0.69$\pm$0.02 &  11.6477$\pm$0.0004  &  0.329$\pm$0.005  &  $\sim$ 5.4915 E+30 \\
54211.48255 &  120$\pm$0.4  &  1245$\pm$0.6  &  1365$\pm$0.8 & 10.38$\pm$0.04 &  112.4808$\pm$0.0009  &  0.496$\pm$0.007  &  $\sim$ 1.0080 E+32 \\
54227.37541 &  15$\pm$0.2  &  75$\pm$0.9  &  90$\pm$0.9  & 5.00$\pm$0.12 &  14.5103$\pm$0.0005  &  0.484$\pm$0.007  &  $\sim$ 6.8411 E+30 \\
54227.49538 &  15$\pm$0.2  &  15$\pm$0.2  &  30$\pm$0.3  & 1.00$\pm$0.02 &  5.0416$\pm$0.0002  &  0.422$\pm$0.006  &  $\sim$ 2.6598 E+30 \\
54227.50538 &  1044$\pm$1.5  &  1260$\pm$2.1  &  2304$\pm$2.6 & 1.21$\pm$0.01 &  133.2189$\pm$0.0013  &  0.258$\pm$0.004  &  $\sim$ 8.1931 E+31 \\
54234.38692 &  364$\pm$0.4  &  741$\pm$0.4 &  1106$\pm$0.5  & 2.03$\pm$0.01 &  27.8203$\pm$0.0010  &  0.277$\pm$0.004  &  $\sim$ 1.3116 E+31 \\
54234.50690 &  373$\pm$0.6  &  787$\pm$0.9  &  1160$\pm$1.1 & 2.11$\pm$0.01 &  164.2169$\pm$0.0027  &  0.488$\pm$0.007  &  $\sim$ 7.7422 E+31 \\
54235.47382 &  1967$\pm$1.1  &  2988$\pm$1.3  &  4955$\pm$1.7 & 1.52$\pm$0.01 &  145.4049$\pm$0.0031  &  0.503$\pm$0.008  &  $\sim$ 1.3288 E+32 \\
54276.37291 &  45$\pm$0.5  &  15$\pm$0.2  &  60$\pm$0.6  & 0.33$\pm$0.01 &  8.5775$\pm$0.0003  &  0.291$\pm$0.004  &  $\sim$ 4.0440 E+30 \\
54276.39774 &  90$\pm$1.1  &  210$\pm$2.5  &  300$\pm$2.7  & 2.33$\pm$0.06 &  51.0005$\pm$0.0018  &  0.342$\pm$0.005  &  $\sim$ 2.4045 E+31 \\
54276.40208 &  15$\pm$0.2  &  60$\pm$0.7  &  75$\pm$0.7  & 4.00$\pm$0.10 &  10.8760$\pm$0.0004  &  0.230$\pm$0.003  &  $\sim$ 5.1277 E+30 \\
54276.40399 &  105$\pm$0.3  &  225$\pm$0.3  &  330$\pm$0.4  & 2.14$\pm$0.01 &  50.4296$\pm$0.0018  &  0.247$\pm$0.004  &  $\sim$ 2.3776 E+31 \\
54276.41007 &  120$\pm$0.2  &  255$\pm$0.4 &  375$\pm$0.4 & 2.13$\pm$0.01 &  65.9255$\pm$0.0013  &  0.256$\pm$0.004  &  $\sim$ 3.1082 E+31 \\
54276.41927 &  30$\pm$0.2  &  30$\pm$0.2  &  60$\pm$0.2 & 1.00$\pm$0.01 &  12.2865$\pm$0.0004  &  0.465$\pm$0.007  &  $\sim$ 5.7927 E+30 \\
54276.43185 &  30$\pm$0.2  &  15$\pm$0.2  &  45$\pm$0.2 & 0.50$\pm$0.01 &  6.6471$\pm$0.0002  &  0.441$\pm$0.007  &  $\sim$ 3.1339 E+30 \\
\hline
\end{tabular}
\end{center}
\end{table*}

\begin{table*}
\begin{center}
\caption{Summary of the reference list and the data used in the analyses.\label{tbl-3}}
\begin{tabular}{lllrlllr}
\hline\hline
Reference 	&	 Observed 	&	Spectral	&	 Number 	&	 Reference 	&	 Observed 	&	Spectral	&	 Number \\
of Data 	&	 Star 	&	Type$^{*}$	&	 of Flare 	&	 of Data 	&	 Star 	&	Type$^{*}$	&	 of Flare \\
\hline
\citet{Mac68} 	&	 AD Leo 	&	M4.5Ve 	&	21	&	\citet{Pan81} 	&	 EV Lac 	&	M4.5V 	&	 11 \\
\citet{Mac68} 	&	 V1054 Oph 	&	M3.5Ve 	&	11	&	\citet{Pan82a} 	&	 EV Lac 	&	M4.5V 	&	 7 \\
\citet{Her69} 	&	 DO Cep 	&	M4.0V	&	10	&	\citet{Pan82b} 	&	 AD Leo 	&	M4.5Ve 	&	 22 \\
\citet{Cri70} 	&	 AD Leo 	&	M4.5Ve 	&	2	&	\citet{Osa73} 	&	 AD Leo 	&	M4.5Ve 	&	 14 \\
\citet{Cri71} 	&	 EV Lac 	&	M4.5V 	&	3	&	\citet{Pan83} 	&	 EV Lac 	&	M4.5V 	&	 12 \\
\citet{Cri72} 	&	 AD Leo 	&	M4.5Ve 	&	1	&	\citet{Her83} 	&	 AD Leo 	&	M4.5Ve 	&	 6 \\
\citet{Cri73a} 	&	 EV Lac 	&	M4.5V 	&	3	&	\citet{Pet84} 	&	 AD Leo 	&	M4.5Ve 	&	 82 \\
\citet{Rod73} 	&	 EV Lac 	&	M4.5V 	&	4	&	\citet{Pan85} 	&	 EV Lac 	&	M4.5V 	&	 20 \\
\citet{Kap73} 	&	 AD Leo 	&	M4.5Ve 	&	2	&	\citet{Pet86} 	&	 AD Leo 	&	M4.5Ve 	&	 99 \\
\citet{Cri73b} 	&	 EV Lac 	&	M4.5V 	&	1	&	\citet{Tsv86a} 	&	 EV Lac 	&	M4.5V 	&	 5 \\
\citet{Mof74} 	&	 Wolf 424 	&	M5	   &	11	&	\citet{Tsv86b} 	&	 EV Lac 	&	M4.5V 	&	 17 \\
\citet{Mof74} 	&	 UV Cet 	&	M6.0V 	&	52	&	\citet{Her87} 	&	 AD Leo 	&	M4.5Ve 	&	 14 \\
\citet{Mof74} 	&	 EV Lac 	&	M4.5V 	&	21	&	\citet{HerA88} 	&	 DO Cep 	&	M4.0V	&	 5 \\
\citet{Mof74} 	&	 EQ Peg 	&	M3.5 	&	42	&	\citet{HerH88} 	&	 AD Leo 	&	M4.5Ve 	&	 6 \\
\citet{Mof74}) 	&	 YZ CMi 	&	M4.5V 	&	51	&	\citet{Pan88} 	&	 EV Lac 	&	M4.5V 	&	 22 \\
\citet{Mof74} 	&	 YY Gem 	&	dM1e 	&	14	&	\citet{Pet91} 	&	 V577 Mon 	&	M4.5V 	&	 72 \\
\citet{Mof74} 	&	 AD Leo 	&	M4.5Ve 	&	8	&	\citet{Ant91} 	&	 AD Leo 	&	M4.5Ve 	&	 4 \\
\citet{Mof74} 	&	 CN Leo 	&	M6.0V 	&	102	&	\citet{Van96} 	&	 YZ CMi 	&	M4.5V 	&	 8 \\
\citet{Osa74a} 	&	 YZ CMi 	&	M4.5V 	&	7	&	\citet{Let97} 	&	 EV Lac 	&	M4.5V 	&	 193 \\
\citet{Kap74} 	&	 AD Leo 	&	M4.5Ve 	&	4	&	\citet{Tov97} 	&	 EV Lac 	&	M4.5V 	&	 20 \\
\citet{Osa74b} 	&	 AD Leo 	&	M4.5Ve 	&	4	&	\citet{Pan00} 	&	 YZ CMi 	&	M4.5V 	&	 7 \\
\citet{Mof75} 	&	 AD Leo 	&	M4.5Ve 	&	27	&	\citet{Dal10} 	&	 EV Lac 	&	M4.5V 	&	 98 \\
\citet{San75} 	&	 YZ CMi 	&	M4.5V 	&	1	&	\citet{Dal10} 	&	 AD Leo 	&	M4.5Ve 	&	 110 \\
\citet{Nic75} 	&	 DO Cep 	&	M4.0V	&	22	&	\citet{Dal10} 	&	 V1054 Oph 	&	M3.5Ve 	&	 40 \\
\citet{Ich78} 	&	 YZ CMi 	&	M4.5V 	&	7	&	\citet{Dal10} 	&	 EQ Peg 	&	M3.5 	&	 73 \\
\citet{Con79} 	&	 EV Lac 	&	M4.5V 	&	23	&	\citet{Dal11a} 	&	 V1005 Ori 	&	M0Ve 	&	 41 \\
\citet{Con80} 	&	 EV Lac 	&	M4.5V 	&	13	&	\citet{Dal11b} 	&	 V1285 Aql 	&	M3.0V 	&	 83 \\
\citet{Avg80} 	&	 EV Lac 	&	M4.5V 	&	5	&	\citet{Dal11c} 	&	 DO Cep 	&	M4.0V	&	 89 \\
\hline
\end{tabular}
\end{center}
$^{*}$ All the spectral types were taken from the SIMBAD database.\\
\end{table*}

\begin{table*}
\begin{center}
\caption{The parameters derived from the OPEA model and the results of the t-Test analyses.\label{tbl-4}}
\begin{tabular}{lr}
\hline\hline
\textbf{Parameter}	&	\textbf{Value} $\pm$\textbf{Error}	\\
\hline
\multicolumn{2}{c}{Parameters of The OPEA Function:} \\
$Plateau$ =	&	2.146$\pm$0.065	\\
$y_{0}$ =	&	0.673$\pm$0.088	\\
$k$ =	&	0.003621$\pm$0.000584	\\
$Span$ =	&	1.473$\pm$0.097	\\
$Half-Life$ =	&	191.40 	\\
$r^{2}$ of the model =	&	0.94 	\\
\multicolumn{2}{c}{The t-Test Analysis:} \\
Mean =	&	2.163$\pm$0.037	\\
Std. Deviation =	&	0.082 	\\
\hline
\end{tabular}
\end{center}
\end{table*}

\begin{table*}
\begin{center}
\caption{The values of the best histograms, which were estimated by SPSS V17.0 and Prism V5.02 software.\label{tbl-5}}
\begin{tabular}{lccc}
\hline\hline
Parameters  &	In This Project	&	Literature	&	Combined	\\
	&	(Figure 5)	&	(Figure 6)	&	(Figure 7)	\\
\hline						
Total Number of Flares =	&	554	&	1118	&	1672	\\
Minimum Value of $\tau_{d}$/$\tau_{r}$ =	&	0.07	&	0.09	&	0.07	\\
25$\%$ Percentile =	&	1.00	&	2.02	&	1.51	\\
Median Value of $\tau_{d}$/$\tau_{r}$ =	&	2.00	&	4.79	&	3.50	\\
75$\%$ Percentile =	&	4.31	&	11.13	&	8.81	\\
Maximum Value of $\tau_{d}$/$\tau_{r}$ =	&	58.33	&	206.39	&	206.39	\\
Mean Value of $\tau_{d}$/$\tau_{r}$ =	&	4.16	&	9.35	&	7.63	\\
Std. Deviation =	&	6.03	&	14.01	&	12.21	\\
Std. Error =	&	0.26	&	0.42	&	0.30	\\
Lower 95$\%$ CI of mean =	&	3.66	&	8.53	&	7.05	\\
Upper 95$\%$ CI of mean =	&	4.66	&	10.18	&	8.22	\\
\hline
\end{tabular}
\end{center}
\end{table*}


\begin{thebibliography}{96}
\bibitem[Abbett et al.(1999)]{Abb99} Abbett, W.P., Hawley, S.L., 1999, ApJ, 521, 906
\bibitem[Allred et al.(2006)]{All06} Allred, J.C., Hawley, S.L., Abbett, W.P., Carlsson, M., 2006, ApJ, 644, 484
\bibitem[Anderson(1979)]{And79} Anderson, C.M., 1979, PASP, 91, 202
\bibitem[Antov et al.(1991)]{Ant91} Antov, A.P., Genkov, V.V., Konstantinova-Antova, R., Kirov, N.K. 1991, IBVS, 3577, 1
\bibitem[Avgoloupis et al.(1980)]{Avg80} Avgoloupis, S., Phylactopoulos, P., Kareklidis, G., Mavridis, L.N., Varvoglis, P., 1980, IBVS, 1793, 1
\bibitem[Benz \& G\"{u}del(2010)]{Ben10} Benz, A.O., G\"{u}del, M., 2010, ARA\&A, 48, 241
\bibitem[Carrington(1859)]{Car59} Carrington, R.C., 1859, MNRAS, 20, 13
\bibitem[Contadakis et al.(1979)]{Con79} Contadakis, M.E., Kareklidis, G., Mavridis, L.N., Tsioumis, A.C., 1979, IBVS, 1653, 1
\bibitem[Contadakis et al.(1980)]{Con80} Contadakis, M.E., Mahmoud, F., Mavridis, L.N., Stavridis, D., 1980, IBVS, 1784, 1
\bibitem[Cristaldi \& Rodon\'{o}(1970)]{Cri70} Cristaldi, S., Rodon\'{o}, M., 1970, A\&AS, 2, 223
\bibitem[Cristaldi \& Rodon\'{o}(1971)]{Cri71} Cristaldi, S., Rodon\'{o}, M., 1971, IBVS, 600, 1
\bibitem[Cristaldi \& Rodon\'{o}(1972)]{Cri72} Cristaldi, S., Rodon\'{o}, M., 1972, IBVS, 682, 1
\bibitem[Cristaldi \& Rodon\'{o}(1973a)]{Cri73a} Cristaldi, S., Rodon\'{o}, M., 1973a, IBVS, 759, 1
\bibitem[Cristaldi \& Rodon\'{o}(1973b)]{Cri73b} Cristaldi, S., Rodon\'{o}, M., 1973b, IBVS, 836, 1
\bibitem[Dal \& Evren(2010)]{Dal10} Dal, H.A., Evren, S., 2010, AJ, 140, 483
\bibitem[Dal \& Evren(2011a)]{Dal11a} Dal, H.A., Evren, S., 2011a, AJ, 141, 33
\bibitem[Dal \& Evren(2011b)]{Dal11b} Dal, H.A., Evren, S., 2011b, PASP, 123, 659
\bibitem[Dal(2011)]{Dal11c} Dal, H.A., 2011, PASA, 28, 365
\bibitem[Dawson \& Trapp(2004)]{Daw04} Dawson, B., Trapp, R.G., 2004, Basic and Clinical Biostatistics (New York: McGraw-Hill), 61
\bibitem[Doyle(1996)]{Doy96} Doyle, J.G., 1996, A\&A, 307, 162
\bibitem[Favata et al.(2005)]{Fav05} Favata, F., Flaccomio, E., Reale, F., Micela, G., Sciortino, S., Shang, H., Stassun, K.G., Feigelson, E.D., 2005, ApJS, 160, 469
\bibitem[Garc\'{i}a-Alvarez et al.(2008)]{Gar08} Garc\'{i}a-Alvarez, D., Drake, J.J., Kashyap, V.L., Lin, L., Ball, B., 2008, ApJ, 679, 1509
\bibitem[Gershberg(1972)]{Ger72} Gershberg, R.E., 1972, Ap\&SS, 19, 75
\bibitem[Gershberg(2005)]{Ger05} Gershberg, R.E., 2005, Solar-type Activity in Main-sequence Stars (New York: Springer), 53
\bibitem[Gershberg et al.(1999)]{Ger99} Gershberg, R.E., Katsova, M.M., Lovkaya, M.N., Terebizh, A.V. and Shakhovskaya, N.I., 1999, A\&AS, 139, 555
\bibitem[Gershberg \& Shakhovskaya(1983)]{Ger83} Gershberg, R.E., Shakhovskaya, N.I., 1983, Astrophys. Space Sci., 95, 235
\bibitem[Green et al.(1999)]{Gre99} Green, S.B., Salkind, N.J., Akey, T. M., 1999, Using SPSS for Windows: Analyzing and Understanding Data (Upper Saddle River, NJ: Prentice Hall), 50
\bibitem[Grinin(1983)]{Gri83} Grinin, V.P., 1983, Activity in Red-dwarf Stars, Proc. Seventy-first Colloq. (Astrophys. Space Sci. Libr. 102; Dordrecht: Reidel), 613
\bibitem[Gurzadian(1988)]{Gur88} Gurzadian, G.A., 1988, ApJ, 332, 183
\bibitem[Jeffries et al.(2011)]{Jef11} Jeffries, R.D., Jackson, R.J., Briggs, K.R., Evans, P.A., Pye, J.P., 2011, MNRAS, 411, 2099
\bibitem[Haisch(1983)]{Hai83} Haisch, B.M., 1983, in Activity in Red-Dwarf Stars, eds. P.B. Byrne, M. Rodon\'{o}, Reidel, Dordrecht, p.255
\bibitem[Hardie(1962)]{Har62} Hardie R.H., 1962, in Astronomical Techniques, ed. W. A. Hiltner (Chicago: Univ. Chicago Press), 178
\bibitem[Hawley \& Fisher(1992)]{Haw92} Hawley, S.L., Fisher, G.H., 1992, ApJS, 78, 565
\bibitem[Hawley et al.(2003)]{Haw03} Hawley, S.L., Allred, J.C., Johns-Krull, C.M., Fisher, G.H., Abbett, W.P., Alekseev, I., Avgoloupis, S.I., Deustua, S.E., Gunn, A., Seiradakis, J.H., Sirk, M.M., Valenti, J.A., 2003, ApJ, 597, 535
\bibitem[Herr \& Brcich(1969)]{Her69} Herr, R.B., Brcich, J.A., 1969, IBVS, 329, 1
\bibitem[Herr \& Caputo(1988)]{HerA88} Herr, R.B., Caputo, F.M., 1988, IBVS, 3243, 1
\bibitem[Herr \& Charache(1988)]{HerH88} Herr, R.B., Charache, D.H., 1988, IBVS, 3270, 1
\bibitem[Herr \& Frank(1983)]{Her83} Herr, R.B., Frank, J.D., 1983, IBVS, 2426, 1
\bibitem[Herr \& Opie(1987)]{Her87} Herr, R.B., Opie, D.B., 1987, IBVS, 3069, 1
\bibitem[Hodgson(1859)]{Hod59} Hodgson, R., 1859, MNRAS, 20, 15
\bibitem[Ichimura \& Shimizu(1978)]{Ich78} Ichimura, K., Shimizu, Y., 1978, IBVS, 1416, 1
\bibitem[Imanishi et al.(2003)]{Ima03} Imanishi, K., Nakajima, H., Tsujimoto, M., Koyama, K., Tsuboi, Y., 2003, PASJ, 55, 653
\bibitem[Kapoor et al.(1973)]{Kap73} Kapoor, R.C., Sanwal, B.B., Sinvhal, S.D., 1973, IBVS, 810, 1
\bibitem[Kapoor \& Sinvhal(1974)]{Kap74} Kapoor, R.C., Sinvhal, S.D., 1974, IBVS, 901, 1
\bibitem[Katsova et al.(1987)]{Kat87} Katsova, M.M., Badalyan, O.G., and Livshits, M.A., 1987, Astron. Zh., 64, 1243
(= Sov. Astron., 31, 652)
\bibitem[Kay et al.(2003)]{Kay03} Kay, H.R.M., Culhane, J.L., Harra, L.K., Matthews, S.A., 2003, AdSpR, 32, 1051
\bibitem[Kirkup \& Frenkel(2006)]{Kir06} Kirkup, L., Frenkel, R.B., 2006, An Introduction to Uncertainty in Measurement, Cambridge University Press
\bibitem[Landolt(1983)]{Lan83} Landolt, A.U., 1983, AJ, 88, 439
\bibitem[Leto et al.(1997)]{Let97} Leto, G., Pagano, I., Buemi, C.S., Rodon\'{o}, M., 1997, A\&A, 327, 1114
\bibitem[MacConnell(1968)]{Mac68} MacConnell, D.J., 1968, ApJ, 153, 313
\bibitem[Mahmoud(1991)]{Mah91} Mahmoud, F.M., 1991, Ap\&SS, 186, 113
\bibitem[Mahmoud(1993)]{Mah93} Mahmoud, F.M., 1993, Ap\&SS, 209, 237
\bibitem[Mei\v{s}tas(2002)]{Mei02} Mei\v{s}tas, E.G., 2002, High-Sped Three-Channel Photometer (HSTCP) User's Guide, To Mol\'{e}tai version (Vilnius, Astronomical Observatory of Vilnius University)
\bibitem[Monagan et al.(2008)]{Mon08} Monagan, M.B., Geddes, K.O., Heal, K.M., Labahn, G., Vorkoetter, S.M., McCarron, J., DeMarco, P., 2008, Maple Introductory Programming Guide (Maplesoft, a division of Waterloo Maple Inc., Canada)
\bibitem[Moffett(1974)]{Mof74} Moffett, T.J., 1974, ApJS, 29, 1
\bibitem[Moffett(1975)]{Mof75} Moffett, T.J., 1975, IBVS, 997, 1
\bibitem[Motulsky(2007)]{Mot07} Motulsky, H., 2007, GraphPad Prism 5: Statistics Guide (San Diego, CA: GraphPad Software Inc. Press), 94
\bibitem[Nicastro(1975)]{Nic75} Nicastro, A.J., 1975, IBVS, 1045, 1
\bibitem[Osawa et al.(1973)]{Osa73} Osawa, K., Ichimura, K., Shimizu, Y., Okada, T., Okida, K., Yutani, M., Koyano, H., 1973, IBVS, 790, 1
\bibitem[Osawa et al.(1974a)]{Osa74a} Osawa, K., Ichimura, K., Okada, T., Okida, K., Yutani, M., Koyano, H., 1974a, IBVS, 876, 1
\bibitem[Osawa et al.(1974b)]{Osa74b} Osawa, K., Ichimura, K., Shimizu, Y., Koyano, H., 1974b, IBVS, 906, 1
\bibitem[Pandey \& Singh(2008)]{Pan08} Pandey, J.C., Singh, K.P., 2008, MNRAS, 387, 1627
\bibitem[Panov et al.(1983)]{Pan83} Panov, K.P., Asteriadis, G., Mavridis, L.N., 1983, IBVS, 2358, 1
\bibitem[Panov et al.(1982b)]{Pan82b} Panov, K.P., Grigorova, M., Tsintsarova, A., 1982b, IBVS, 2220, 1
\bibitem[Panov et al.(1982a)]{Pan82a} Panov, K.P., Pamukchiev, I., Christov, P., Asteriadis, G., Mavridis, L. N., 1982a, IBVS, 2128, 1
\bibitem[Panov et al.(1985)]{Pan85} Panov, K.P., Piirola, V., Korhonen, T., 1985, IBVS, 2826, 1
\bibitem[Panov et al.(1988)]{Pan88} Panov, K.P., Piirola, V., Korhonen, T., 1988, A\&AS, 75, 53
\bibitem[Panov \& Tsvetkov(1981)]{Pan81} Panov, K.P., Tsvetkov, M.K., 1981, IBVS, 1971, 1
\bibitem[Panov et al.(2000)]{Pan00} Panov, K., Goranova, Yu., Genkov, V., 2000, IBVS, 4917, 1
\bibitem[Petit(1957)]{Pet57} Petit, M., 1957, SvA, 1783
\bibitem[Pettersen et al.(1984)]{Pet84} Pettersen, B.R., Coleman, L.A., Evans, D.S., 1984, ApJ, 282, 214
\bibitem[Pettersen et al.(1986)]{Pet86} Pettersen, B.R., Panov, K.P., Ivanova, M.S., Sandmann, W.H., 1986, A\&AS, 66, 235
\bibitem[Pettersen \& Sundland(1991)]{Pet91} Pettersen, B.R., Sundland, S.R., 1991, A\&AS, 87, 303
\bibitem[Reale et al.(1997)]{Rea97} Reale, F., Betta, R., Peres, G., Serio, S., McTiernan, J., 1997, A\&A, 325, 782
\bibitem[Reale \& Micela(1998)]{Rea98} Reale, F., Micela, G., 1998, A\&A, 334, 1028
\bibitem[Reale et al.(1988)]{Rea88} Reale, F., Peres, G., Serio, S., Rosner, R., Schimitt, J.H.M.M., 1988, ApJ, 328, 256
\bibitem[Reeves \& Warren(2002)]{Ree02} Reeves, K.K., Warren, H.P., 2002, ApJ, 578, 590
\bibitem[Rodon\'{o}(1973)]{Rod73} Rodon\'{o}, M., 1973, IBVS, 802, 1
\bibitem[Sanwal(1975)]{San75} Sanwal, B.B., 1975, IBVS, 998, 1
\bibitem[Shibata \& Yokoyama(1999)]{Shi99} Shibata, K., Yokoyama, T., 1999, ApJ, 526, L49
\bibitem[Shibata \& Yokoyama(2002)]{Shi02} Shibata, K., Yokoyama, T., 2002, ApJ, 577, 422
\bibitem[Skumanich \& McGregor(1986)]{Sku86} Skumanich, A., McGregor, K., 1986, Adv. Space Phys., 6, 151
\bibitem[Spanier \& Oldham(1987)]{Spa87} Spanier, J., Oldham, K.B., 1987, An Atlas of Function (Washington, DC: Hemisphere Publishing Corporation Press), 233
\bibitem[Stauffer \& Hartmann(1986)]{Sta86} Stauffer, J.R., Hartmann, L.W., 1986, ApJS, 61, 531.
\bibitem[Tamazian et al.(2008)]{Tam08} Tamazian, V.S., Docobo, J.A., Balega, Y.Y., Melikian, N.D., Maximov, A.F., Malogolovets, E.V., 2008, AJ, 136, 974
\bibitem[Temmer et al.(2001)]{Tem01} Temmer, M., Veronig, A., Hanslmeier, A., Otruba, W., and Messerotti, M., 2001, A\&A 375, 1049
\bibitem[Tovmassian et al.(1997)]{Tov97} Tovmassian, H.M., Recillas, E., Cardona, O., Zalinian, V.P., 1997, RMxAA, 33, 107
\bibitem[Tsvetkov et al.(1986b)]{Tsv86b} Tsvetkov, M.K., Antov, A.P., Tsvetkova, A.G., 1986b, IBVS, 2972, 1
\bibitem[Tsvetkov et al.(1986a)]{Tsv86a} Tsvetkov, M.K., Tsvetkova, K.P., Melikian, N.D., 1986a, IBVS, 2954, 1
\bibitem[van den Oord \& Barstow(1988)]{VanB88} van den Oord, G.H.J., Barstow, M.A., 1988, A\&A, 207, 89
\bibitem[van den Oord et al.(1996)]{Van96} van den Oord, G.H.J., Doyle, J.G., Rodon\'{o}, M., Gary, D.E., Henry, G.W., Byrne, P.B., Linsky, J.L., Haisch, B.M., Pagano, I., Leto, G., 1996, A\&A, 310, 908
\bibitem[van den Oord et al.(1988)]{VanA88} van den Oord, G.H.J., Mewe, R., Brinkman, A.C., 1988, A\&A, 205, 181
\bibitem[Veeder(1974)]{Vee74} Veeder, G.J., 1974, AJ, 79, 702V
\bibitem[Vilhu et al.(1986)]{Vil86} Vilhu, O., Neff, J.E., Walter, F.M., 1986, in ESA SP-263, New Insight in Astrophysics, ed. E. J. Rolfe (Noordwijk: ESA), 113
\bibitem[Wall \& Jenkins(2003)]{Wal03} Wall, J.W. \& Jenkins, C.R., 2003, In Practical Statistics For Astronomers, Cambridge University Press, p.79
\bibitem[Yamamoto et al.(2002)]{Yam02} Yamamoto, T.T., Shiota, D., Sakajiri, T., Akiyama, S., Isobe, H., Shibata, K., 2002, ApJ, 579, L45
\bibitem[Yokoyama \& Shibata(1998)]{Yok98} Yokoyama, T., Shibata, K., 1998, ApJ, 494, L113
\end{thebibliography}
\end{document}